\documentclass[letterpaper, 10 pt, conference]{ieeeconf}  

\IEEEoverridecommandlockouts                              
\usepackage{graphics} 
\usepackage{epsfig} 
\usepackage{amsmath} 
\usepackage{amssymb}  

\usepackage[ruled]{algorithm2e}
\usepackage{amsmath}

\renewcommand{\baselinestretch}{0.99}

\newtheorem{theorem}{Theorem}

\newtheorem{lemma}{Lemma}
\newtheorem{definition}{Definition}

\newtheorem{corollary}{Corollary}

\title{\LARGE \bf
A Structured Systems Approach for Optimal Actuator-Sensor Placement in  Linear Time-Invariant Systems}

\author{S\'ergio Pequito$^{1,2}$ $\quad$ Soummya Kar$^{1}$ $\quad$ A. Pedro Aguiar$^{2}$
\thanks{Partially supported by grant SFRH/BD/33779/2009, from Funda\c{c}\~ao para a Ci\^encia e a Tecnologia and  the CMU-Portugal (ICTI) program}
\thanks{
$^{1}$Dept. of Electrical and Computer Engineering, Carnegie Mellon University, Pittsburgh, PA 15213.}
\thanks{
$^{2}$Dept. of Electrical and Computer Engineering, Institute for System and Robotics, Instituto Superior T\'ecnico, Technical University of Lisbon, Lisbon, Portugal.
        }%
}

\begin{document}

\maketitle
\thispagestyle{empty}
\pagestyle{empty}

\begin{abstract}

In this paper we address the actuator/sensor allocation problem for linear time invariant (LTI) systems. Given the structure of an autonomous linear dynamical system, the goal is to design the structure of the input matrix (commonly denoted by $B$) such that the system is structurally controllable with the restriction that each input be  dedicated, i.e., it can only control directly a single state variable. We provide a methodology that addresses this design question: specifically, we determine the minimum number of dedicated inputs required to ensure such structural controllability, and characterize, and characterizes  all  (when not unique) possible configurations of the \emph{minimal} input matrix $B$. Furthermore, we show that the proposed solution methodology incurs \emph{polynomial complexity} in the number of state variables. By duality, the solution methodology may be readily extended to the structural design of the corresponding minimal output matrix (commonly denoted by $C$) that ensures structural observability.

\end{abstract}

\section{INTRODUCTION}

This paper is motivated by the dearth of scalable techniques for the analysis and synthesis of different large-scale complex systems, notably ones which tackle design and decision making in a single framework. Examples include power systems, public or business organizations, large manufacturing systems, wireless control systems, biological complex networks, and formation control, to name a few. Focusing on the last case, consider, for instance, the synchronization problem in vehicular formations: Given a communication topology for inter-vehicle information exchange and the individual vehicle (agent) models, we are often interested in addressing the following questions:
\begin{itemize}
\item What is the smallest subset of agents (and specifically which ones), that need a \emph{dedicated input} (i.e., a control that directly affects a single state variable), such that the system is controllable?
\item Similarly, what is the smallest subset of agents (and specifically which ones) that need to be equipped with a \emph{dedicated output} (i.e., an output that measures directly  a single state variable), such that the entire network state may be estimated?
\end{itemize}

Referring to the vehicular-formation scenario, the different agents that require dedicated controls to achieve system controllability, are those that play the role of \emph{leaders}. Under infrastructure and operational  constraints, identifying the smallest subset of such agents clearly maximizes the efficiency of the system. The concerns posed above go beyond the vehicular-formation example and are applicable to wider classes of large-scale multi-agent scenarios.

To address these problems, we will resort to structural systems theory~\cite{dionSurvey}, in which the main idea is to reformulate and  study of an equivalent class of systems for which system-theoretic properties are investigated based on only the sparsity pattern (i.e., the location of zeroes and non-zeroes) of the state space representation matrices. Such an approach is particularly helpful when dealing with systems parameter uncertainties. Analysis using structural systems provides  system-theoretic guarantees that hold for almost all values of the parameters, except for a manifold of zero Lebesgue measure \cite{Reinschke:1988}, which may further be characterized algebraically~\cite{SSCrevisited}.  Properties such as controllability and observability are referred to as  \emph{structural controllability}\footnote{A pair $(A_0,B_0)$ is structurally controllable if and only if $\forall\epsilon>0$, there exist completely controllable pairs $(A_1,B_1)$ and $(A_{2},B_{2})$ with the same structure (location of zeroes and non-zeroes) as $(A_0,B_0)$, such that $\|A_1-A_0\|>\epsilon$, $\|B_1-B_0\|>\epsilon$, and $\|A_{2}-A_{0}\|<\epsilon$, $\|B_{2}-B_{0}\|<\epsilon$.} and \emph{structural observability}  in this framework, as they hold in general, i.e, for almost all non-zero entries in  the  state space representation. 
With this, our design objective may be precisely formulated as follows:

\vspace{-0.1cm}

\subsection*{Problem Statement}

Given
\vspace{-0.1cm}
\begin{equation}
\dot x=\bar A x,
\label{eqA}
\end{equation}
\vspace{-0.5cm}

\noindent where $\bar A$ represents the structural pattern of $A$ (i.e., the locations of zeroes and non-zeroes only), our goal is 
\begin{itemize}
\item[${\mathcal{P}}_1$]  Design $\bar B$ (i.e., find the structural pattern)  with a minimum number of dedicated inputs, such that $(\bar A,\bar B)$ is structurally controllable. Stated formally, characterize (all) $\overline{B}\in\mathbb{R}^{n\times p}$ such that 
\[
\forall_{j\in \{1,\cdots, p\}}\ \exists^1_{i\in\{1,\cdots,n\}} \ \bar{B}_{ij}\neq 0,
\]
 in other words
\begin{equation}
\dot  x=\bar Ax +\sum_{j=1}^p \bar{b}_{i_j}u_{j}, \ i_j\in\{1,\cdots,n\},
\label{systemStruct}
\end{equation}
where $\bar b_{i_j}$ represents the $i_j$-th canonical vector and $u_j\in \mathbb{R}$ represents the  $j$-th control, such that system \eqref{systemStruct} is structurally  controllable, $p\in \mathbb{N}$, and there exists no other $p'<p$ that satisfies the previous requirement.
\end{itemize}

Solution of $\mathcal{P}_{1}$ also addresses the corresponding optimal (minimal placement) structural observability output matrix design problem by invoking the duality between estimation and control in LTI systems.
\hfill $\diamond$
\vspace{-0.07cm}

The literature on structured systems theory is extensive; see [2,4-6]\nocite{Lin_1974,largeScale,Reinschke:1988,Murota:2009:MMS:1822520} for earlier work, see also \cite{dionSurvey} for a recent survey. For applications  to optimal sensor and actuator placements, the reader may refer to  \cite{reviewPlacement} and references therein; however, these approaches mostly lead to combinatorial implementation complexity in the number of state vertices (agents), or are often based on simplified heuristic-based reductions of the optimal design problems. Systematic approaches to structured systems based design were investigated recently in the context of different application scenarios, see, for example, [8-13]\nocite{Khan,Khan2,Sundaram,DBLP:journals/automatica/BoukhobzaH11a,Bullo,liu11}; for instance,  in network estimation, as in [8,11] \nocite{Khan,DBLP:journals/automatica/BoukhobzaH11a},  where strategies for output (sensor) placement are provided, ensuring only sufficient (but not necessarily minimal) conditions for structural observability, whereas in [8,13]\nocite{Khan, Bullo} applications to power system state estimation are explored. From the structural observability viewpoint, as a key contrast to the above approaches, we study the constrained output placement problem, specifically, in which the sensors are dedicated, in that, they may only measure a single state variable. The formulation that is closest to our setup in terms of minimal actuator placement arises  from biological complex networks~\cite{liu11}, where the concept of a \textit{driving vertex} is introduced, i.e., a state vertex through which a subset of  the graph vertices can be controlled. One of the problems addressed in that work may be stated as follows: What is the minimum number of driving vertices that make the entire network controllable? However, the results in \cite{liu11} hold only for the case in which the  system graph is strongly connected.  In contrast, in this paper, in addition to providing the  minimum number of dedicated inputs (driving vertices) for  generic system matrices (digraphs), we also characterize the set of all possible minimal feasible configurations, i.e., allocations with the minimum number of dedicated inputs that ensure structural controllability.

\vspace{-0.06cm}

To summarize, in terms of the dedicated input placement problem, the  main contributions of this paper include: 1) the  identification of the minimum number of dedicated inputs (to ensure structural controllability); 2)   characterization of  all such feasible minimal dedicated input placement configurations  (i.e.,  the design of $\bar B$ up to column permutations); 3) algorithmic generation of a minimal feasible input configuration in polynomial complexity (in the number of state variables).

\vspace{-0.06cm}


The rest of the this paper is organized as follows. Section \ref{prelim} reviews some concepts and introduces results (some of them new) in structural systems theory and establish their relations to graph-theoretic constructs. Subsequently, in Section \ref{mainresults} we present the main technical results (proof outlines being relegated to the appendices), followed  by an illustrative example in Section \ref{illustrativeexample}. Conclusions are presented in Section \ref{conclusions}. Finally, Section \ref{conclusions} concludes the paper and discusses avenues for further research.

\section{PRELIMINARIES AND TERMINOLOGY}\label{prelim}

In this section we recall some classical concepts in structural systems, introduced in \cite{Lin_1974}. 

Given a dynamical system \eqref{eqA}, an efficient approach to the analysis of its structural properties  is to associate it with a directed graph (digraph) $\mathcal{D}=(V,E)$,  in which $V$ denotes a set of \textit{vertices} and $E$ represents a set of \textit{edges}, such that, an edge $(v_j,v_i)$ is directed from vertex $v_j$ to vertex $v_i$. Denote by $\mathcal{X}=\{x_1,\cdots,x_n\}$ and $\mathcal{U}=\{u_1,\cdots,u_p\}$ the set of state vertices and input vertices, respectively. Denote by $ \mathcal{E}_{\mathcal{X},\mathcal{X}}=\{(x_i,x_j):\ [\bar A]_{ij}\neq 0\}$ and $\mathcal{E}_{\mathcal{U},\mathcal{X}}=\{(u_j,x_i):\ [\bar B]_{ij}\neq 0\}$, 
to define $\mathcal{D}(\bar A)=(\mathcal{X},\mathcal{E}_{\mathcal{X},\mathcal{X}})$ and $\mathcal{D}(\bar A,\bar B)=(\mathcal{X}\cup \mathcal{U},\mathcal{E}_{\mathcal{X},\mathcal{X}}\cup \mathcal{E}_{\mathcal{U},\mathcal{X}} )$. A digraph $\mathcal{D}_s=(V_s,E_s)$ with $V_s\subset V$ and $E_s\subset E$ is called a \textit{subgraph} of $\mathcal{D}$. If $V_s=V$, $\mathcal{D}_s$ is said to \textit{span} $\mathcal{D}$. A sequence of edges $\{(v_1,v_2),(v_2,v_3),\cdots,(v_{k-1},v_k)\}$, in which all the vertices are distinct, is called \textit{an  elementary path} from $v_1$ to $v_k$. When $v_k$ coincides with $v_1$, the sequence is called a \textit{cycle}.

In addition, we will require the following graph theoretic notions~\cite{Cormen:2001:IA:580470}: A digraph $\mathcal{D}$ is said to be strongly connected if there exists a directed path between any two pairs of vertices. A strongly connected component (SCC) is a maximal subgraph $\mathcal{D}_S=(V_S,E_S)$ of $\mathcal{D}$ such that for every $v,w \in V_S$ there exists a path from $v$ to $w$ and from $w$ to $v$.
Note that, an SCC may have several paths between two vertices and the path from $v$ to $w$ may comprise some vertices not in the path from $w$ to $v$.  Visualizing each SCC as a virtual node (or supernode), one may generate a \textit{directed acyclic graph} (DAG), in which each node corresponds to a single SCC and a directed edge exists between two SCCs \emph{iff} there exists a directed edge connecting the corresponding SCCs in the original digraph. The DAG associated with $\mathcal{D}=(V,E)$ may be efficiently generated in $\mathcal{O}(|V|+|E|)$~\cite{Cormen:2001:IA:580470}, where  $|V|$ and $|E|$  denote the number of vertices in $V$ and the number of edges in $E$, respectively. The SCCs in a DAG can be characterized as follows

\begin{definition}\label{linkedSCC}
An SCC is said to be linked  if it has at least one incoming/outgoing edge from another SCC. In particular,  an SCC is  \textit{non-top linked} if it has no incoming edges to its vertices from the vertices of another SCC and \textit{non bottom linked} if it has no outgoing edges to another SCC.
\hfill $\square$
\end{definition}

For any two vertex sets $S_{1}, S_{2}\subset V$, we define the   \textit{bipartite graph} $\mathcal{B}(S_1,S_2,E_{S_1,S_2})$ associated with $D=(V,E)$, to be a directed graph (bipartite), whose vertex set is given by $S_{1}\cup S_{2}$ and the edge set $E_{S_1,S_2}$ by
$
E_{S_1,S_2}=\{(s_1,s_2)\in E \ :\ s_1 \in S_1, s_2 \in S_2  \ \}.
$

Given $\mathcal{B}(S_1,S_2,E_{S_1,S_2})$, a matching $M$ corresponds to a subset of edges in $E_{S_1,S_2}$ that do not share vertices, i.e., given edges  $e=(s_1,s_2)$ and $e'=(s_1',s_2')$ with $s_1,s_1' \in S_1$ and $s_2,s_2'\in S_2$, $e, e' \in M$ only if $s_1\neq s_1'$ and $s_2\neq s_2'$. A maximum matching $M^{\ast}$ may then be defined as a matching $M$ that has the largest number of edges among all possible matchings. The maximum matching problem may be solved  efficiently in $\mathcal{O}(\sqrt{|S_1\cup S_2|}|E_{S_1,S_2}|)$ \cite{Cormen:2001:IA:580470}. Vertices in $S_1$ and $S_2$ are \textit{matched vertices} if they belong to an edge in the maximum matching $M^*$, otherwise, we designate the vertices as \textit{unmatched vertices}. If there are no unmatched vertices, we say that we have a \textit{perfect match}. It is to be noted  that a maximum matching $M^*$ may not be unique. 

For ease of referencing, in the sequel, the term \emph{right-unmatched vertices} (w.r.t. $\mathcal{B}(S_1,S_2,E_{S_1,S_2})$ and a maximum matching $M^{\ast}$) will refer to only those vertices in $S_{2}$ that do not belong to a matched edge in $M^{\ast}$.

\vspace{-0.15cm}

\subsection{Structural Systems}
\vspace{-0.15cm}

Given a digraph, we further define the following special subgraphs:\\
\textbf{-} Consider $\mathcal{D}(\bar A)$ : \emph{State Stem} - An elementary path composed exclusively by state vertices, or a single state vertex.\\
\textbf{-} Consider $\mathcal{D}(\bar A,\bar B)$ : \emph{Input Stem} - An elementary path composed of an input vertex (the root) linked to the root of a state stem.\\
\textbf{-} Consider $\mathcal{D}(\bar A)$ : \emph{State Cactus} -  Defined recursively  as follows: A state stem is a state cactus.  A state cactus connected  to a cycle from any point other than the tip is also a state cactus.\\
\textbf{-} Consider $\mathcal{D}(\bar A,\bar B)$ : \emph{Input Cactus} - Defined recursively  as follows: An input stem with at least one state vertex is an input cactus. An input cactus connected  to a cycle from any point other than the tip is also an input cactus.\\
\textbf{-} Consider $\mathcal{D}(\bar A)$ : \emph{Chain} - A group of disjoint cycles (composed by state vertices) connected to each other in a sequence, or a single cycle.\\
\indent  The root and the tip of a stem are also the root and tip of the associated cactus.

Furthermore, recall the following  result:

\begin{theorem}[\cite{Shields_Pearson:1976}]\label{teo1}
For an LTI system $\dot{x}=Ax+Bu$, the following statements are equivalent:
\begin{itemize}
\item[i)] The corresponding structured linear system $(\bar A,\bar B)$ is structurally controllable.
\item[ii)] The digraph $\mathcal{D}(\bar A,\bar B)$ is spanned by a disjoint union of input cacti.
\hfill$\square$
\end{itemize}
\end{theorem}

Note that , by definition, an input cactus may have an input vertex linked to several state vertices, i.e., the input vertex may connect to the root of a state stem (i.e., input stem) and could be linked to one or more states in a chain. 


\vspace{-0.17cm}

\subsection{Relation between Maximum Matching and Concepts in Structural Systems}
\vspace{-0.1cm}

The following results provide a bridge between structural systems concepts and graph constructs such as maximum matching. These results will be used to characterize the minimal dedicated input configurations in Section III. The proofs are based on standard graph theoretic properties and relegated to the appendix.

\begin{lemma}\label{cycleMatch}
 Let $\mathcal{D}(\bar A)=(\mathcal{X},\mathcal{E}_{\mathcal{X},\mathcal{X}})$ and $M^*$ a maximum matching associated with $\mathcal{B}(\mathcal{X},\mathcal{X},\mathcal{E}_{\mathcal{X},\mathcal{X}})$. Then, if $M^*$ is a perfect match, the edges in $M^*$ correspond to a disjoint union of cycles in $\mathcal{D}(\bar A)$.
\hfill $\square$
\end{lemma}

\begin{lemma}[Maximum Matching Decomposition]\label{MMD}
Let $\mathcal{D}(\bar A)=(\mathcal{X},\mathcal{E}_{\mathcal{X},\mathcal{X}})$ and $M^*$ a maximum matching associated with $\mathcal{B}(\mathcal{X},\mathcal{X},\mathcal{E}_{\mathcal{X},\mathcal{X}})$. Suppose $M^{\ast}$ consists of a non-empty set of right-unmatched vertices. Then, $M^*$ (i.e., the edges in $M^{\ast}$), together with the set of isolated  vertices, constitutes a disjoint union of cycles and  state stems (with roots in the right-unmatched vertices) that  span $\mathcal{D}(\bar A)$. Moreover, such a decomposition is minimal, i.e., no other spanning subgraph decomposition of $\mathcal{D}(\bar{A})$ into state stems and cycles contains strictly fewer number of state stems. 
\hfill $\square$
\end{lemma}

In other words, the maximum matching problem leads to two different kinds of matched edge sequences in $M^*$; sequences of edges in $M^*$ starting in  right-unmatched state vertices,   and the remaining sequences of edges that start and end in a matched vertices. These sequences represent state stems and cycles, respectively. 

In case a graph is composed of multiple SCCs, we define
\begin{definition}\label{originateMM}
Let $\mathcal{D}(\bar A)=(\mathcal{X},\mathcal{E}_{\mathcal{X},\mathcal{X}})$ and $M^*$ a maximum matching associated with $\mathcal{B}(\mathcal{X},\mathcal{X},\mathcal{E}_{\mathcal{X},\mathcal{X}})$. We say that a subgraph of $\mathcal{D}(\bar A)$, denoted by $\mathcal{S}=(\mathcal{X}^S,\mathcal{E}_{\mathcal{X}^S,\mathcal{X}^S})$, \emph{originates a perfect match with respect to } $M^*$ if 
$
M'=\{e: \ e\in M^* \text{ and } e \in \mathcal{E}_{\mathcal{X}^S,\mathcal{X}^S}\}
$
is such that $|M'|=|\mathcal{X}^S|$, i.e, in other words, there is a subset of edges  $M'\subset M^*$, such that  $M'$ is a maximum matching to $\mathcal{B}(\mathcal{X}^S,\mathcal{X}^S,\mathcal{E}_{\mathcal{X}^S,\mathcal{X}^S})$. 
\hfill $\square$
\end{definition}

\begin{definition}\label{topassignedSCC}
Let $\mathcal{D}(\bar A)=(\mathcal{X},\mathcal{E}_{\mathcal{X},\mathcal{X}})$ and $M^*$ a maximum matching associated with $\mathcal{B}(\mathcal{X},\mathcal{X},\mathcal{E}_{\mathcal{X},\mathcal{X}})$. 
A non-top linked SCC is said to be a \emph{top assignable SCC} if it contains at least one right-unmatched vertex (with respect to $M^*$).
\hfill $\square$
\end{definition}

Note that the total number of top assignable SCCs may depend on the particular maximum matching $M^{\ast}$ (not unique in general) under consideration; as such we may define:

\begin{definition}
Let $\mathcal{D}(\bar A)=(\mathcal{X},\mathcal{E}_{\mathcal{X},\mathcal{X}})$ and $M^*$ a maximum matching associated with $\mathcal{B}(\mathcal{X},\mathcal{X},\mathcal{E}_{\mathcal{X},\mathcal{X}})$. 
The \emph{maximum assignability  index} of $\mathcal{B}(\mathcal{X},\mathcal{X},\mathcal{E}_{\mathcal{X},\mathcal{X}})$ is the maximum number of top assignable SCCs that a maximum matching $M^*$ may lead to.
\hfill $\square$
\end{definition}

\vspace{-0.3cm}

\section{MAIN RESULTS}\label{mainresults}

In this section we present the main results of this paper (an outline of the proofs is relegated to the appendix), broadly centered on the following two issues:
\begin{itemize}
\item Determine the minimum number of dedicated inputs to be allocated to ensure structural controllability.
\item Describe the set of all possible \emph{minimal feasible input configurations} (i.e., allocation configurations with the minimum number of dedicated inputs) which lead to structural controllability.
\end{itemize}

The first result relates the minimum number of dedicated inputs necessary to ensure structural controllability to the structure of the cacti associated with the system digraph. Specifically, we have the following:

\begin{theorem}\label{theorem2}
Given the system (state) digraph $\mathcal{D}(\bar A)$, the minimum number of dedicated inputs required to ensure structural controllability is equal to the minimum number of disjoint state cacti that span $\mathcal{D}(\bar A)$.
\hfill $\square$
\end{theorem}

Theorem \ref{theorem2} reduces the problem of finding the minimum number of dedicated inputs to that of finding the minimum number of disjoint state cacti spanning $\mathcal{D}(A)$. The next set of results are concerned with explicitly characterizing the latter number by invoking the relationship (see, for instance, Lemma 2) between cacti decompositions and more readily computable graph constructs such as maximum matching.

\vspace{-0.2cm}

\subsection*{Minimum Number of Dedicated Inputs}

The following characterization of the minimum number of disjoint state cacti spanning $\mathcal{D}(\bar{A})$ (and, hence, the minimum number of dedicated inputs holds:

\begin{theorem}[Minimum Number of Dedicated Inputs]

Let $\mathcal{D}(\bar A)=(\mathcal{X},\mathcal{E}_{\mathcal{X},\mathcal{X}})$, where $|\mathcal{X}|=n$ and $M^*$ be a  maximum matching associated with $\mathcal{B}(\mathcal{X},\mathcal{X},\mathcal{E}_{\mathcal{X},\mathcal{X}})$. Moreover, let $\mathcal{S}^i=(\mathcal{X}^i,\mathcal{E}_{\mathcal{X}^i,\mathcal{X}^i})$, for $i=1,\cdots,k$, be the non-top linked SCCs of the DAG representation of $\mathcal{D}(\bar A)$, with $\mathcal{X}^i\subset \mathcal{X}$ and $\mathcal{E}_{\mathcal{X}^i,\mathcal{X}^i}\subset \mathcal{E}_{\mathcal{X},\mathcal{X}}$. Then, the minimum number of dedicated inputs $p$ is given by
\vspace{-0.3cm}
\begin{equation}
p=m+\beta-\alpha,
\label{eq2}
\end{equation}
\vspace{-0.15cm}
where
\begin{itemize}
\item[-] $m=|\mathcal{V}|$ and $\mathcal{V}$ denotes the set of right-unmatched vertices with respect to $M^*$;
\item[-] $\beta$ is the number of non-top linked SCCs;
\item[-] $\alpha$ is the maximum assignability  index of $\mathcal{B}(\mathcal{X},\mathcal{X},\mathcal{E}_{\mathcal{X},\mathcal{X}})$.
\hfill $\square$
\end{itemize}
\label{mainThm2}
\end{theorem}

First note that, in Theorem \ref{mainThm2}, the number of right-unmatched vertices (and hence, the minimum number of dedicated inputs $p$) does not depend on the specific instantiation of the maximum matching $M^*$ being considered (which is not unique in general). Moreover, it may be readily verified from the definitions, that if $\mathcal{D}(\bar A)$ is strongly connected,  we have $\beta=1$,in Theorem 3, and $\alpha$ may only assume two values, $0$ or $1$, depending on whether $m=0$ or $m=1$ respectively. As such, Theorem 3 may be simplified significantly if $\mathcal{D}(A)$ is known to be strongly connected:

\begin{corollary}
Let $\mathcal{D}(\bar A)=(\mathcal{X},\mathcal{E}_{\mathcal{X},\mathcal{X}})$ be strongly connected with  $|\mathcal{X}|=n$. Let $M^{\ast}$ be a maximum matching associated with $\mathcal{B}(\mathcal{X},\mathcal{X},\mathcal{E}_{\mathcal{X},\mathcal{X}})$ and designate by $\mathcal{V}$ the set of right-unmatched vertices. Then, the number of dedicated inputs  $p$
\begin{enumerate}
\item[-]  is $1$ if  $|\mathcal{V}|=0$ (i.e., there is a perfect match),
\item[-]  is $m$ if $|\mathcal{V}|=m>0$.
\hfill $\square$
\end{enumerate}
\label{mainThm1}
\end{corollary}

Characterizations of the required number of minimal dedicated inputs similar to Corollary 1 were presented in \cite{liu11}; however, the strong connectivity of the system digraph is necessary for the validity of Corollary 3. In general, when $\mathcal{D}(\bar{A})$ consists of multiple SCCs, the reduction of Theorem 3 to Corollary 1 is not valid (as illustrated by the example design scenario in Section IV).

Theorem \ref{mainThm2} provides the minimum number of required  dedicated inputs, hence the minimum number of columns in $\bar B$ (each with only one non-zero entry) required to ensure structural controllability. We now explicitly characterize all such $\bar{B}$'s   (up to a permutation of the columns). Each such combination will be referred to as a \emph{minimal feasible input configuration}.
\vspace{-0.1cm}
\subsection*{Minimal Feasible Input Configurations}
\vspace{-0.1cm}

A minimal feasible input configuration will be denoted by a $p$-tuple $<x_{i_{1}},\cdots,x_{i_{p}}>$ of states, where $p$ corresponds to the minimal number of dedicated inputs ensuring structural controllability (see~\eqref{eq2} in Theorem 3) and $i_{k}\in\{1,\cdots,n\}$ for all $k=1,\cdots,p$. In other words, a $p$-tuple  $<x_{i_{1}},\cdots,x_{i_{p}}>$ corresponds to a minimal feasible input configuration if allocating dedicated actuators (inputs) to each of the $p$ states $x_{i_{k}}$ in the tuple leads to structural controllability. Also, note that each such $p$-tuple corresponds to a unique canonical $\bar{B}$; hence, identifying the set of all possible canonical minimal $\bar{B}$'s is equivalent to identifying all such $p$-tuples of minimal feasible input configurations.

Also, denote by $\Theta$ the set of all possible minimal feasible input configurations, i.e.,
\vspace{-0.2cm}
\begin{align*}
\Theta=&\{< x_{i_1},\cdots, x_{i_p}> : x_{i_1}\neq\cdots \neq x_{i_p} \text{ and if a } \\
&\text{ dedicated input is assigned to each } x_{i_{k}}, \text{where}\\
& \ i_{k}\in\{1,\cdots,n\} \text{ and } k=1,\cdots,p, \text{ the resulting} \\ 
&\text{  LTI system is structurally controllable}\}.
\end{align*}
\vspace{-0.5cm}

 Note that, by the above definition, a minimal feasible input configuration is invariant to any permutation of the states in its associated $p$-tuple representation; a permutation of the states in the $p$-tuple leads to the same dedicated input assignment.  The following set of results concerns the efficient description and enumeration of the set $\Theta$ of all possible minimal feasible input configurations. The key driving factor behind an efficient representation of $\Theta$ is the existence of subsets $\Theta^{j}\subset\mathcal{X}$, $j=1,\cdots,p$, such that $\Theta$ is \emph{almost} (to be made precise soon) the Cartesian product of the $\Theta^{j}$'s, i.e., $\Theta\simeq\Theta^{1}\times\cdots\Theta^{p}$, i.e., it will be shown that, up to permutation and some \emph{natural constraints} on the $\Theta^{j}$'s, a $p$-tuple $<x_{i_{1}},\cdots,x_{i_{p}}>$ belongs to $\Theta$ \emph{iff} $x_{i_{j}}\in\Theta^{j}$ for all $j=1,\cdots,p$. Specifically, we have the following:
\begin{theorem}[Naturally Constrained Partitions]\label{mainThm3}
Let $\mathcal{D}(\bar A)=(\mathcal{X},\mathcal{E}_{\mathcal{X},\mathcal{X}})$, with $|\mathcal{X}|=n$ and $M^*$ denoting  a  maximum matching associated with $\mathcal{B}(\mathcal{X},\mathcal{X},\mathcal{E}_{\mathcal{X},\mathcal{X}})$, where $\mathcal{V}=\{v_1,\cdots,v_m\}$ be the set of $m$ right-unmatched vertices. Moreover, let $\mathcal{S}^l=(\mathcal{X}^l,\mathcal{E}_{\mathcal{X}^l,\mathcal{X}^l})$, for $l=\{1,\cdots,k\}$, be the non-top linked SCCs of the DAG representation of $\mathcal{D}(\bar A)$, with $\mathcal{X}^l\subset \mathcal{X}$ and $\mathcal{E}_{\mathcal{X}^l,\mathcal{X}^l}\subset \mathcal{E}_{\mathcal{X},\mathcal{X}}$. The set $\Theta$ of minimal feasible input configurations may be characterized as follows: 

There exists $\Theta^j \subset \mathcal{X}$, $j=1,\cdots,p$ (where $p$ denotes the minimum number of dedicated inputs as in  Theorem \ref{mainThm2}), given by
\begin{itemize}
\item[-] $j=1,\cdots,m$
\vspace{-0.2cm}
\begin{align*}
 \Theta^j=&\{x : \ \left(\mathcal{V}-\{v_j\} \right) \cup \{x\} \text{ for } x \in \mathcal{X} \text{ is the set of }\\
&\qquad \text{ right-unmatched vertices for some  }\\
&\qquad \text{ maximum matching of } \mathcal{B}(\mathcal{X},\mathcal{X},\mathcal{E}_{\mathcal{X},\mathcal{X}})\},
\end{align*}
\item[-]  $j=m+1,\cdots,p: \   \Theta^j=\bigcup\limits_{l\in\{1,\cdots,k\}}\mathcal{S}^l,$
\end{itemize}
such that
\vspace{-0.15cm}
\begin{equation}
< x_{i_1},\cdots, x_{i_p}>\ \in \Theta 
\label{feasibleConfig}
\end{equation}
\emph{ if and only if} the following hold:
\begin{itemize}
\item[i)] $x_{i_j}\in\Theta^j$ $i_{j}\in\{1,\cdots,n\}$ for all $j=1,\cdots,p$ and $i_{j}\neq i_{j^{\prime}}$ for $j\neq j^{\prime}$,  
\item[ii)] $\ x_{i_j} \in \Theta^j$ and $x_{i_{j'}} \in \Theta^{j'}$ for $j\neq j^{\prime}$ implies that $x_{i_j}$ and $x_{i_{j'}}$  are the root to two different minimal state stems,
\item[iii)] for each non-top linked SCC $\mathcal{S}^{l}$, there is at least one state variable in $<x_{i_{1}},\cdots,x_{i_{p}}>$ that belongs to $\mathcal{S}^{l}$.
\vspace{-0.15cm}

\hfill $\square$
\end{itemize}
\end{theorem}
\vspace{-0.1cm}
Note that the sets $\Theta^{j}$ are defined on the basis of the specific maximum matching $M^{\ast}$ in consideration. However, as will be evident from the proofs, up to a permutation of the indices $j$, $j=1,\cdots,p$, the $\Theta^{j}$'s are independent of the actual instantiation of $M^{\ast}$ (which may not be unique).
We refer to the sets $\Theta^j$, for $j=1,\cdots,p $ endowed with the \emph{natural constraints} as the \emph{natural constrained partitions} of $\Theta$.
Given the description in Theorem \ref{mainThm3}, we are interested in understanding the computational (algorithmic) complexity of implementing the \emph{natural constrained partitions}, as well as understanding how to use such characterization to compute iteratively a minimal feasible input configuration. This is the scope of the next result, stated as follows:

\begin{theorem}[Complexity]\label{NCPtheorem}
 
Let the hypotheses of Theorem 4 hold. Then, there exist algorithms of polynomial complexity (in the number of state vertices) to implement the following procedures:
\begin{itemize}
\item[1)] obtaining the minimum number of dedicated inputs;
\item[2)] constructing the natural constrained partitions, $\Theta^{j}$'s, of $\Theta$;
\item[3)] generating a minimal feasible input configuration iteratively.
\hfill $\square$
\end{itemize}
\end{theorem}


First, note that, although by Theorem 5, there exist polynomial algorithms to constructing the $\Theta^{j}$'s and obtaining a minimal feasible input configuration, listing specifically all possible minimal configurations may be combinatorial (the number of such configurations could be exponential in the number of state vertices).  The  algorithmic procedures to obtaining the minimum number of dedicated inputs and the natural constrained partitions are provided in Appendix B. An example presented in the following section.

\vspace{-0.1cm}

\section{AN ILLUSTRATIVE EXAMPLE}\label{illustrativeexample}

\vspace{-0.2cm}

The following example  illustrates the procedure to obtaining the minimum number of dedicated inputs. Consider  a set of $6$ agents that do share information. Let us  assume that each agent $i$ has a predefined path $y_i:\mathbb{R}\rightarrow \mathbb{R}^N$, to follow  parametrized by a time-dependent parameter $\gamma_i : \mathbb{R} \rightarrow \mathbb{R}$, such that $y_i(\gamma_i (t))$ provides the position of an agent $i$ at time instant $t$.  Suppose we are interested in addressing the controlled synchronization problem, which consists of some predefined parameter specification $\gamma^*\in \mathbb{R}^6$.

Moreover, suppose that the autonomous system may be described as follows:
\vspace{-0.2cm}
\begin{align}
\dot \gamma_1&=-k_1\gamma_1,  \ \dot \gamma_2=-k_2\gamma_2,  \  \dot \gamma_5=-k_9\gamma_4, \ \dot \gamma_6=-k_{10}\gamma_4,\notag\\ 
\dot \gamma_3&=-k_3\gamma_1-k_4\gamma_2 -k_5\gamma_4, \
\dot \gamma_4=-k_6\gamma_3-k_7\gamma_5-k_8\gamma_6, \label{eqdyacop}
\end{align}
\vspace{-0.4cm}

 \noindent where $k_i \in \mathbb{R}^+$, $i=1,\cdots,10$ are prescribed gains. 

The question that we address is  the following: what is the minimum subset of agents that we need to assign a dedicated control such that any specified $\gamma^*$ is achievable in finite time?

\begin{figure}[htb]
\centering
\includegraphics[scale=0.23]{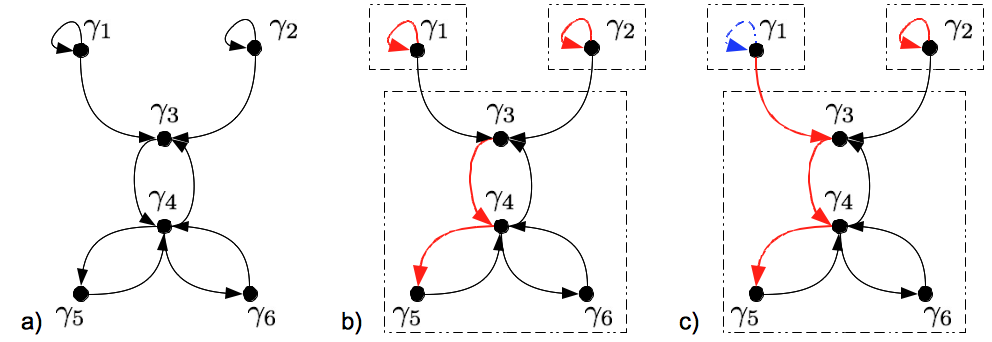}
\caption{ a) The digraph $\mathcal{D}(A)$ modeling the inter-dependecy of the parameters of the $6$ agents;  b) The SCCs are depicted  by the rectangles and the red edges represent a maximum matching (recall Lemma \ref{MMD}); c) Removing the edge in blue   from $\mathcal{D}(\bar A)$ forces $\gamma_1$ to be a right-unmatched vertex.}
\label{fig:max1}
\end{figure}

From \eqref{eqdyacop} we obtain   the structured (autonomous) system

\begin{equation}
\dot \gamma=\underbrace{\left[ \begin{array}{cccccc}
\times&0&0&0&0&0\\
0&\times&0&0&0&0\\
\times &\times&0&\times &0&0\\
0&0&\times&0&\times&\times\\
0&0&0&\times&0&0\\
0&0&0&\times&0&0\\
\end{array}
\right]}_{\bar A} \gamma
\label{exempleSync}
\end{equation}
where $\gamma=[\gamma_1 \ \gamma_2 \ \gamma_3 \ \gamma_4\ \gamma_5]^T$ and $\times$ denote the non-zero entries. System \eqref{exempleSync} provides the digraph representation $\mathcal{D}(\bar A)=(\mathcal{X},\mathcal{E}_{\mathcal{X},\mathcal{X}})$ as indicated in Fig. \ref{fig:max1}-a). 

The question that we address is the following: what is the minimum subset of agents to which we need to assign dedicated controls such that any specified  is attainable in finite time?

To this end, we consider the following steps:

\noindent \textbf{Step 1} Compute a maximum matching $M^*$ associated with $\mathcal{B}(\mathcal{X},\mathcal{X},\mathcal{E}_{\mathcal{X},\mathcal{X}})$.  Figure \ref{fig:max1}-b) represents in red, the edges belonging to  the maximum matching $M^*$ (note that $M^{\ast}$ is not unique in general).

\begin{figure}[htb]
\centering
\includegraphics[scale=0.23]{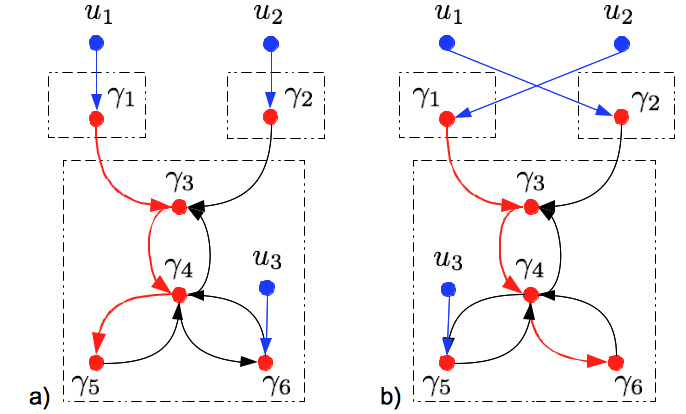}
\caption{Digraph $\mathcal{D}(\bar A)$  where the SCCs are depicted by rectangles. The red edges and state vertices identify the state stems and the blue vertices correspond to the dedicated inputs connected to the roots of the state stems. In a) we have the feasible minimum input configuration $<\gamma_1,\gamma_2,\gamma_6>$  and in b) the feasible minimum input configuration $<\gamma_2,\gamma_1,\gamma_5>$.}
\label{fig:max2}
\end{figure}

\noindent \textbf{Step 2} Find the minimum number of dedicated inputs required to ensure structural controllability. To this end, given $M^*$, the set of right-unmatched vertices are $\mathcal{V}=\{\gamma_3,\gamma_6\}$, and, hence, in  Theorem \ref{NCPtheorem} we have $m=2$  and $\beta=2$ corresponding to the non-top linked SCCs. To find $\alpha$ in  Theorem \ref{NCPtheorem}, consider the alternatives for $\gamma_3$, that are in the non-top linked SCCs. For instance, let us verify if $\gamma_1$ is a possible alternative to $\gamma_3$. To this end, we force $\gamma_1$ to be a right-unmatched vertex, by removing all incoming edges to $\gamma_1$ on the original digraph (represented in blue in Fig. \ref{fig:max2}) and compute a new maximum matching. This new maximum matching is depicted in Figure \ref{fig:max1}-c). Since, the new maximum matching consists of the same number of edges, $\gamma_1$ is a possible alternative and the SCC containing $\gamma_1$ is assignable.  Now, consider $\gamma_6$ and let's verify if there exists an alternative in the corresponding non-top linked SCC. Recall that $\gamma_1$ is fixed because it's in an assignable non-top linked SCC and our goal is to find the maximum assignability index. Let  us consider that $\gamma_2$ is also fixed, (by removing its self loop), and compute a new maximum matching. This provides $\{\gamma_1,\gamma_2,\gamma_3\}$ as right-unmatched vertices, which implies that the maximum matching has one less edge with respect to the previous. Since there is only one non-top linked assignable SCC, in Theorem  \ref{NCPtheorem} we have $\alpha=1$. Hence, we need $p=m+\beta-\alpha=2+2-1=3$ dedicated inputs.

\noindent \textbf{Step 3}  In this step, we characterize all possible feasible minimal input configurations (up to permutation) by resorting to Theorem \ref{mainThm3}. The generic algorithm (Algorithm 3) is presented in Appendix B.  We start by using Algorithm 3-(1) (see Appendix B) as a subroutine to compute the set of possible alternatives to each right-unmatched vertex. Initially set $\Theta^1=\{\gamma_1\}$, $\Theta^2=\{\gamma_6\}$. In order to extend these sets, let's verify which alternatives to $\gamma_1$ provide  maximum matchings with the same number of edges as $M^*$. Consider the original $\mathcal{B}(\mathcal{X},\mathcal{X},\mathcal{E}_{\mathcal{X},\mathcal{X}})$  and fix $\gamma_6$ (we are just exploring alternatives to $\gamma_1$), by removing the edges in $\mathcal{E}_{\mathcal{X},\mathcal{X}}$ that end in $\gamma_6$. Moreover, to explore if the remaining vertices $\{\gamma_2,\gamma_3,\gamma_4,\gamma_5\}$ are viable alternatives, consider for $i=2,3,4,5$, $\mathcal{B}^{\gamma_i,\gamma_6}(\mathcal{X},\mathcal{X},\mathcal{E}_{\mathcal{X},\mathcal{X}}-\{(.,\gamma_6)-\{(.,\gamma_i)\}\}), \text{ and } M^{\gamma_i,\gamma_6}$ 
the corresponding (arbitrary) maximum matching. It turns out that $|M^{\gamma_2,\gamma_6}|=|M^{\gamma_3,\gamma_6}|=|M^{\gamma_5,\gamma_6}|=4$ and $|M^{\gamma_3,\gamma_6}|=3$, hence $\Theta^1=\{\gamma_1,\gamma_2,\gamma_3,\gamma_5\}$. Similarly, by fixing $\gamma_1$ we have $\Theta^2=\{\gamma_6,\gamma_5\}$. Because $\beta-\alpha=1$ in Theorem \ref{mainThm2}, by Theorem \ref{mainThm3}  we need an extra partition given by $\Theta^3=\{\gamma_1,\gamma_2\}$. Together with the natural constraints we have the characterization of all minimal feasible input configurations.

\noindent \textbf{Step 4} In this step we create iteratively the feasible configurations from the natural constrained partitions in Step 3).  The natural constraints impose that first we assign a dedicated input to at least one state variable to $\alpha$ non-top linked assignable SCC. Thus,
\begin{enumerate}
\item Picking $\gamma_1$ from $\Theta^1$, followed by $\gamma_2$ from $\Theta^3$ leaves us with the choice of $\gamma_5,\gamma_6$ from $\Theta^2$, leading to the minimal feasible configurations   $<\gamma_1,\gamma_2,\gamma_5>$ and $<\gamma_1,\gamma_2,\gamma_6>$, respectively.  Equivalently, the above correspond to following structures of the matrix $\bar B$ (up to column permutations):
\vspace{-0.15cm}
\[
\bar B=\overbrace{\begin{pmatrix} \times & 0 &0\\ 0&\times&0\\0&0&0\\0&0&0\\0&0&\times\\0&0&0\end{pmatrix}}^{<\gamma_1,\gamma_2,\gamma_5>}, 
\bar B= \overbrace{\begin{pmatrix} \times & 0 &0\\ 0&\times&0\\0&0&0\\0&0&0\\0&0&0\\0&0&\times\end{pmatrix}}^{<\gamma_1,\gamma_2,\gamma_6>}.
\]
\vspace{-0.15cm}

\item Similarly, picking $\gamma_1$ from $\Theta^3$, followed by $\gamma_2$ from $\Theta^1$ leaves us with  the choice of $\gamma_5,\gamma_6$ from $\Theta^2$, hence providing the same configurations as in (1).
\end{enumerate}

\section{CONCLUSIONS AND FURTHER RESEARCH}\label{conclusions}

 In this paper we provided a systematic method with polynomial complexity (in the number of the state variables) to obtain the minimum number of dedicated inputs (through the structural design of B up to column permutation), and characterize all possible solutions that ensures structural controllability of a given LTI system. By duality, the results extend to the corresponding structural observability output design. A natural extension of the current framework consists of obtaining minimal allocations for general cost constrained placement problems, in which actuator-sensor placements may incur different costs at different state vertices. This problem is more challenging; a natural way to proceed is to modify the constructions of the natural constrained partitions suitably so as to incorporate the non-homogeneous assignment costs.
\vspace{-0.3cm}





\small

\section*{APPENDIX A}

\subsection*{Proof of Lemma \ref{cycleMatch}} 

Both sets in the bipartite graph have the same number $n$ of vertices, and assume that $\mathcal{X}=\{x^1,\cdots, x^n\}$. Thus, $|M^*|=n$ and we may only have the following two possibilities:

\begin{itemize}
\item $(x^i,x^i)\in M^*$ \textit{which is a self-loop (i.e, a cycle)}
\item \textit{for $(x^j,x^k)\in M^*$ there is a sequence of edges such that the last edge in $M^*$ is given by $(., x^j)$}, otherwise, $x^j$ is a right-unmatched vertex and we have a contradiction with the fact that $M^*$ is a perfect match.
\end{itemize}
The assertion follows immediately by noting that each of the two possibilities above lead to a cycle in $\mathcal{D}(\bar A)$ (by properly sequencing the edges in $M^{\ast}$) and that the resulting cycles are disjoint in $\mathcal{D}(\bar A)$ as the associated edges correspond to a matching.
\hfill $\blacksquare$
\vspace{-0.1cm}

\subsection*{Proof of Lemma \ref{MMD}} 

That $M^{\ast}$ together with the set of isolated digraph vertices span $\mathcal{D}(\bar{A})$, may be readily verified from the definition of maximum matching, see, for example, [15]. 

In the following we prove the minimality of the decomposition (through state-stems and cycles) achieved by $M^{\ast}$. To this end, note that the following generic digraph properties may be verified from the definitions:

(1) Let $\mathcal{C}$ denote an arbitrary spanning decomposition of $\mathcal{D}(\bar{A})$ into disjoint subgraphs  of state stems and cycles. Then, the edges in $\mathcal{C}$ define a matching $M^{\mathcal{C}}$ (not necessarily maximum) for $\mathcal{B}(\mathcal{X},\mathcal{X},\mathcal{E}_{\mathcal{X},\mathcal{X}})$.

(2) The root of a state stem belonging to a matching $M$ of $\mathcal{B}(\mathcal{X},\mathcal{X},\mathcal{E}_{\mathcal{X},\mathcal{X}})$ is necessarily a right-unmatched vertex with respect to $M$.

With the above properties, we now establish the desired minimality of $M^{\ast}$ by contradiction. Assume, on the contrary, there exists a spanning decomposition $\mathcal{C}$ of $\mathcal{D}(\bar{A})$ into disjoint subgraphs  of state stems and cycles, such that, $\mathcal{C}$ consists of strictly fewer state stems than $M^{\ast}$. Then, by property (1) above, the corresponding matching $M^{\mathcal{C}}$ consists of fewer state stems than $M^{\ast}$. Since, by property (2) above, each state-stem in $M^{\mathcal{C}}$ corresponds to a unique right-unmatched vertex (and similarly for $M^{\ast}$), we conclude that $M^{\mathcal{C}}$ consists of strictly fewer right-unmatched vertices than $M^{\ast}$. This clearly contradicts the fact, that, $M^{\ast}$ is a maximum matching. Hence, the desired assertion follows.

\vspace{-0.15cm}

\subsection*{Proof of Theorem \ref{theorem2}} 

The proof follows by contradiction. Let $N$ and $M$  denote the minimum number of dedicated inputs and state cacti that span $\mathcal{D}(\bar A)$, respectively and assume that $M\neq N$. Thus, we have two possibilities $M>N$ or $M<N$. 

If $M>N$ then there is at least one state cacti that, in particular, has no dedicated input in its origin. Hence, denoting by $\bar B$ the canonical input matrix resulting from the assumed minimal placement configuration, we note that the augmented digraph $\mathcal{D}(\bar A,\bar B)$ is not spanned by a disjoint union of input cacti. Indeed, this follows from the fact that a state cactus with a dedicated input is in particular an input cactus.

Now consider the possibility $M<N$. Similarly, denoting by $\bar B$ the (canonical) input matrix, the hypothesis implies that $\mathcal{D}(\bar A,\bar B)$ is spanned by $M$ disjoint input cacti, which, together with the restriction that we have dedicated inputs and the characterization in Theorem \ref{teo1} further imply that $N$ is not the minimum number of dedicated inputs. Thus, the claim follows by contradiction.\hfill $\blacksquare$

\vspace{-0.15cm}

\subsection*{Proof of Theorem \ref{mainThm2}} 

The proof is lengthy. We sketch the main outline below, the intermediate technical details being omitted. 

In order to establish the desired assertion, first note that, by Theorem 2 it suffices to show that the minimal number of disjoint state cacti that span $\mathcal{D}(\bar{A})$ is $m+\beta-\alpha$. To this end, let $\mathcal{T}$ denote an arbitrary decomposition of $\mathcal{D}(\bar{A})$ into disjoint state cacti and $\mathbb{T}$ denote the set of all such $\mathcal{T}$. Clearly, we are interested in obtaining $\min_{\mathcal{T}\in\mathbb{T}}|\mathcal{T}|$, where $|\mathcal{T}|$ denotes the number of cacti in $\mathcal{T}$. 

By Lemma 2, each maximum matching of $\mathcal{B}(\mathcal{X},\mathcal{X},E_{\mathcal{X},\mathcal{X}})$ generates a $\mathcal{T}\in\mathbb{T}$ and let $\mathbb{T}_{M}\subset\mathbb{T}$ be the subset consisting of such $\mathcal{T}$'s. Noting the minimality  of maximum matching based decompositions (again by Lemma 2), it may be argued (we omit some techincal details here) that the desired minimation may be reduced to that over the set $\mathbb{T}_{M}$, i.e., $p=\min_{\mathcal{T}\in\mathbb{T}_
{M}}|\mathcal{T}|$. 

Noting that each maximum matching consists of $m$ state stems, each element $\mathcal{T}\in\mathbb{T}_{M}$ contains at least $m$ disjoint state cacti. Hence, $|\mathcal{T}|\geq m$. Hence $p\geq m$, i.e., one needs to assign at least $m$ dedicated inputs. Now fixing a $\mathcal{T}\in\mathbb{T}_{M}$ again (and let $M^{\ast}$ denote the associated maximum matching), 
suppose there are  $c(M^*)$ non-top linked SCCs that originate perfect matches with respect to $M^*$ and which do not contain right-unmatched vertices of $M^*$. Note that, by definition of $\alpha$, it follows that $c(M^*)\ge \beta-\alpha$. Clearly,  for each non-top linked SCC that originate a perfect matches with respect to $M^*$, we clearly  need to assign an additional  dedicated input to one of its vertices, because, by definition of non-top linked SCC, there is no incoming edge to the states that it consists of.  
Hence, $p=\min_{\mathcal{T}\in\mathbb{T}_{M}}|T|=m+\min_{M^*}c(M^*)\ge m+ \beta-\alpha$. Now note that, by definition of maximum assignability index, there exists a maximum matching $(M^*)'$ such that $c((M^*)')=\beta-\alpha$. Hence, it follows that $p=m+\beta-\alpha$. \hfill $\blacksquare$

\vspace{-0.25cm}

\subsection*{Proof of Theorem \ref{mainThm3}} 

\noindent $[\Rightarrow]$ First, note that i) holds by definition of $\Theta^i$. From the proof of Theorem \ref{mainThm2} it may readily be seen that the natural constraint iii) must hold, otherwise the tuple is not a minimal feasible input configuration. Secondly, natural constraint ii) holds, otherwise, it increases the number of dedicated inputs required to have a minimal feasible input configuration. \\
$[\Leftarrow]$ Let $<x_{i_{1}},\cdots,x_{i_{p}}>$ be a $p$-tuple for which the natural constraints hold. Note that, by the hypotheses of Theorem 4, the set of $p$-tuples and the natural constraints are defined w.r.t. a given maximum matching $M^{\ast}$.  From the proof of Theorem \ref{mainThm2}, we note that $m$ entries in the $p$-tuple correspond to right-unmatched vertices. In particular, the natural constraints imply that out of these $m$ variables, $\alpha$ are in non-top linked SCCs, and the remaining $\beta-\alpha$  arbitrary variables from each of those  non-top linked SCCs that do not contain any of the previous $\alpha$ right-unmatched vertices. Hence, the $p$-tuple is a  minimal feasible input configuration. \hfill $\blacksquare$
\vspace{-0.2cm}

\section*{APPENDIX B}
\vspace{-0.08cm}

In this appendix we introduce the algorithmic procedures used to compute: 1) the minimum number of dedicated inputs; 2) the natural constrained partitions, and 3) an iterative method to compute a minimal feasible configuration. Using these results we then prove Theorem \ref{NCPtheorem} (at the end of this section). In the following we introduce two strategies that work as subroutines for the main algorithms: STRATEGY A as a subroutine to compute the naturally constrained partitions of $\Theta$ and the natural constraints i)-ii) introduced in Theorem \ref{mainThm3}; STRATEGY B as a subroutine to compute the maximum assignablity index and the natural constraint iii). Let $\mathcal{B}(S_1,S_2,E_{S_1,S_2})$ be a bipartite graph and $M^*$ an associated maximum matching $M^*$(computed, for instance, using  the Hopcroft-Karp algorithm \cite{Cormen:2001:IA:580470}). Now, we show how to compute a maximum matching such that a specific vertex is right-unmatched (STRATEGY A), or, how to compute a maximum matching with a specific matched edge in it (STRATEGY B).
\vspace{-0.15cm}
\subsection*{STRATEGY A - Find a maximum matching with some vertex as a right-unmatched vertex}\vspace{-0.08cm}

By definition, a right-unmatched vertex  is a vertex that does not belong to the tip of a directed edge from $S_1$ to $S_2$. To this end, let $\mathcal{B}(S_1,S_2,E_{S_1,S_2}-\{(s_1,v) : \ s_1\in S_1\})$be the graph (bipartite) obtained by removing all the edges ending in $v$; then any maximum matching $M^{V}$ of the latter graph contains $v$ as a right-unmatched vertex. Note that in our case (recall Lemma \ref{MMD}), each minimal state stem has its root in a right-unmatched vertex, hence we are interested in the cases where $|M^V|=|M^*|$, in other words, when the number of state stems does not increase.

\vspace{-0.2cm}
\subsection*{STRATEGY B - Find a maximum matching with a given specific edge being a matched edge }

To find a maximum matching for which a prescribed edge $e=(e_1,e_2)$ is  a matched edge, let $M^e=\{e\}\cup M'$, in which $M'$ corresponds to a maximum matching associated with
$\mathcal{B}(S_1-\{e_1\},S_2-\{e_2\},E_{S_1,S_2}-\{(e_1,v_2) : \ v_2 \in S_2\} - \{(v_1,e_2) : \ v_1 \in S_1\})$. It may be seen that $M^{e}$ is in fact a maximum matching and contains $e$.
\hfill $\diamond$

\subsection*{Finding the minimum number of dedicated inputs}

Recall Theorem \ref{mainThm2}. First compute a maximum matching $M^*$ associated with $\mathcal{B}(\mathcal{X},\mathcal{X},\mathcal{E}_{\mathcal{X},\mathcal{X}})$; we then obtain a set of $m$ right-unmatched vertices $\mathcal{V}=\{v_1,\cdots,v_m\}$. Secondly, compute the DAG of $\mathcal{D}(\bar A)$ and denote by $\mathcal{S}^i$, $i=1,\cdots, \beta$  the non-top linked SCCs (and $\beta$ the same as in Theorem \ref{mainThm2}). Now, sequentially try to find a maximum matching such that the right-unmatched vertices are in the non-top linked SCCs and  keep  record to which non-top linked SCC they can be assigned. Based on this record,  the maximum assignability index ($\alpha$ in Theorem \ref{mainThm2}) may be obtained as shown in the following.

\begin{algorithm}[t]
\SetAlgoNoEnd
\small
\KwIn{Set of right-unmatched vertices $\mathcal{V}$, $\mathcal{B}(\mathcal{X},\mathcal{X},\mathcal{E}_{\mathcal{X},\mathcal{X}})$ and $\mathcal{S}^i,\ i=1,\cdots \beta$.}
\KwOut{Set $V$ of right-unmatched vertices in the non-top linked SCCs}
$V=\{\}$;

\For{each $z \in \mathcal{V}$ }{
\For{each $v \in \mathcal{S}^1\cup \cdots\cup \mathcal{S}^{\beta}-V$}{Compute a maximum matching $M^{V}_v$ associated with $\mathcal{B}(\mathcal{X},\mathcal{X},\mathcal{E}_{\mathcal{X},\mathcal{X}}-\{(x,v) : \ x \in \mathcal{X}\} -\{(x,w): \ x \in \mathcal{X}, w \in \mathcal{V}-\{z\} \});$

\If{$|M^{V}_v|=m$ }{$V=V\cup \{v\}$ and EXIT FOR}
}
}
\caption{Find the right-unmatched vertices that belong to non-top linked SCCs}
\label{alg:one}
\end{algorithm}

Algorithm 1 provides us with the right-unmatched vertices that can be in the non-top linked SCCs. Now we need to maximize the number of right-unmatched vertices in different non-top linked SCCs. This procedure, in turn, will be used to provide us with the maximum assignability index. To this end, consider $\mathcal{B}(I_V,I_S,E_{I_V,I_S})$ where
\begin{itemize}
\item $I_V=\{1,\cdots,|V|\}$ is the set of indices of the vertices in $V=\{v_1,\cdots,v_{|V|}\}$, i.e., the right-unmatched vertices that are in the non-top linked SCCs (provided by Algorithm 1);
\item $I_S=\{1,\cdots,\beta \}$ denotes the indices of the non-top linked SCCs;
\item $E_{I_V,I_S}$ is the set of edges $e=(i,j)$, $i\in I_V,\ j\in I_S $ that codifies the possibility of a vertex $v_i$ belonging to a non-top linked SCCs $\mathcal{S}^j$(provided by Algorithm 2).
\end{itemize}
By definition, the maximum assignability index, may  now be readily computed by obtaining a maximum matching associated with $\mathcal{B}(I_V,I_S,E_{I_V,I_S})$.  Finally, the minimum number of dedicated inputs is obtained as $p=m+\beta-\alpha$ by Theorem 3.
\vspace{-0.2cm}

\begin{algorithm}[t]
\SetAlgoNoEnd
\small

\KwIn{ $V$ from Algorithm 1 and $\mathcal{B}(\mathcal{X},\mathcal{X},\mathcal{E}_{\mathcal{X},\mathcal{X}})$ and $\mathcal{S}^j,\ j=1,\cdots,\beta$ }
\KwOut{$E_{I_V,I_S}$ (possible assignments between dedicated input and non-top linked SCCs) }
$E_{I_V,I_S}=\{\}$;

\For{ each $v \in V=\{v_1,\cdots,v_{|V|}\}$ }{
\For{ each $w \in \mathcal{S}^1\cup \cdots\cup \mathcal{S}^{\beta}- (V-\{v\})$ }{
Compute a maximum matching  $M$ associated with  $\mathcal{B}(\mathcal{X},\mathcal{X},\mathcal{E}_{\mathcal{X},\mathcal{X}}-\{(x,w): \ x \in \mathcal{X}\}-\{(x,v_1) : \ x \in \mathcal{X}, v_1 \in (V-\{v\})\} );$

\If{$|M|=m$}{ (consider $v\equiv v_i$)

\For{ $j=1,\cdots,\beta$ }{
\If{$w \in \mathcal{S}^j$}{$ E_{I_V,I_S}=E_{I_V,I_S}\cup\{(i,j)\}$ }
}

 }

}

}

\caption{Determine $E_{I_V,I_S}$}
\label{alg:two}
\end{algorithm}

\subsection*{Generating a minimal feasible input configuration}

Given a maximum matching $M^{\ast}$ of $\mathcal{D}(\bar{A})$, in this section we provide an iterative algorithm to generate a minimal feasible input configuration, which is a tuple $<x_{i_1},\cdots,x_{i_p}>$ invariant to permutation. Broadly, the idea is to iteratively choose a variable from the natural constrained partitions (note that we may select the first variable from any of the partitions) and update the natural constraints based on which a new variable may be chosen. 

Specifically, the strategy to create a minimal feasible input configuration consists of the  following: first, chose the variables that are in the non-top linked SCCs  that ensure the maximum number
of top assignable SCCs. Secondly,  select one state variable for each non-top linked SCC that was not selected in step one. Finally, the third step consists of selecting the variables corresponding to right-unmatched vertices with respect to $M^{\ast}$, that were not  used in step one.

\begin{algorithm}[t]
\SetAlgoNoEnd
\small
\KwIn{The set of right-unmatched vertices $V=\{v_1,\cdots,v_m\}$ and $\mathcal{B}(\mathcal{X},\mathcal{X},\mathcal{E}_{\mathcal{X},\mathcal{X}})$}
\KwOut{Feasible minimum input configuration}

$configuration=\{\}$;

$assignments=\{\}$;
 \%pairs of the form (dedicated input index, non-top linked SCC index)

$\mathcal{I}=\{1,\cdots,m\}$;

$\mathcal{J}=\{1,\cdots,\beta\}$;\\

\% $assignments[i], i=1,2$ represents the set comprising the first and second entry of the pairs in $assignments$;

\For{$1:\alpha$}{

(1) Compute $\Theta^i$ for $i\in\mathcal{I}-assignments[1]$ considering $\mathcal{D}=(\mathcal{X},\mathcal{E}_{\mathcal{X},\mathcal{X}}-\{(x,v): x \in \mathcal{X}, v \in configuration\})$ [Use strategy A and definition of $\Theta^i$ in Theorem \ref{mainThm3}];
\vspace{0.2cm}

\For{$x \in \Theta^i\cap\mathcal{S}^j$ such that  \\ \qquad\quad $(i,j)\in \mathcal{I}\times\mathcal{J}-assignments$}{
(2) Verify which variables $x$ provide a maximum assignability index, if not remove $x$ from $ \Theta^i$

[Use STRATEGY B applied to $E_{I_V,I_S}$ using $assignments \cup \{(i,j)\}$];
}
\vspace{0.2cm}

(3) Select a state variable $y$ form $\mathcal{S}^j$, where  $(i,j)\in \mathcal{I}\times\mathcal{J}-assignments$;

$configuration=configuration\cup \{y\}$;

$assignments=assignments\cup \{(i,j)\}$;
}

\vspace{0.2cm}

\For{$\alpha+1:\beta$}{
(4) Select a state variable   $y$ form $\mathcal{S}^j$, where  $j\in \mathcal{J}-assignments[2]$;
}
\vspace{0.2cm}

\For{ $\beta+1:p$}{
(5) Compute $\Theta^i$ for $i\in\mathcal{I}-assignments[1]$ considering $\mathcal{D}=(\mathcal{X},\mathcal{E}_{\mathcal{X},\mathcal{X}}-\{(x,v): x \in \mathcal{X}, v \in configuration\})$;
}

\caption{Create iteratively a feasible minimum input configuration}
\label{alg:three}
\end{algorithm}

\vspace{-0.3cm}

\subsection*{Proof of Theorem \ref{NCPtheorem}} 

\noindent $1)$ \emph{Minimum number of dedicated inputs}: It consists of computing an initial maximum matching in $\mathcal{O}(\sqrt{|\mathcal{X}|}|\mathcal{E}_{\mathcal{X},\mathcal{X}}|)$~\cite{Cormen:2001:IA:580470} to find $m$. To determine the DAG (may be implemented by executing \emph{Depth-First Search} twice, with complexity $\mathcal{O}(|\mathcal{X}|+|\mathcal{E}_{\mathcal{X},\mathcal{X}}|)$~\cite{Cormen:2001:IA:580470}). The maximum assignability index is computed by executing Algorithm 1 and Algorithm 2; the first algorithm has complexity  $m\bar N\mathcal{O}(\sqrt{|\mathcal{X}|}|\mathcal{E}_{\mathcal{X},\mathcal{X}}|)$, with $\bar N$ ($\le N$) being the number of state vertices in the non-top linked SCCs. Algorithm 2 has complexity $\mathcal{O}(\bar N |V|)$ with $|V|$ being at most $m$.\\
$3)$ Step (1) (Algorithm 3) executes the maximum matching at most $\alpha p |\mathcal{X}|$ times, hence its complexity is $\mathcal{O}(\alpha p |\mathcal{X}|\sqrt{|\mathcal{X}|}|\mathcal{E}_{\mathcal{X},\mathcal{X}}|)$. Step (2) (Algorithm 3)
 uses at most $p|\mathcal{X}| \bar N$ executions of the Algorithm 2. Step (3) (Algorithm 3) consists of  a choice of a single state variable by the designer, hence of constant complexity. Step (4) (Algorithm 3) consists of $\beta-\alpha$ executions of a constant complexity procedure, hence, is of constant complexity again. Finally, Step (5) (Algorithm 3) executes $p-\beta$ times a procedure of complexity $\mathcal{O}(p|\mathcal{X}|\sqrt{|\mathcal{X}|}|\mathcal{E}_{\mathcal{X},\mathcal{X}}|)$.\\
$2)$ The natural constrained partitions can be computed by using single execution of Algorithm 3. \hfill $\blacksquare$


\vspace{-0.2cm}

\small
\bibliographystyle{IEEEtran}
\bibliography{IEEEabrv,acc2013}

\begin{thebibliography}{10}
\providecommand{\url}[1]{#1}
\csname url@samestyle\endcsname
\providecommand{\newblock}{\relax}
\providecommand{\bibinfo}[2]{#2}
\providecommand{\BIBentrySTDinterwordspacing}{\spaceskip=0pt\relax}
\providecommand{\BIBentryALTinterwordstretchfactor}{4}
\providecommand{\BIBentryALTinterwordspacing}{\spaceskip=\fontdimen2\font plus
\BIBentryALTinterwordstretchfactor\fontdimen3\font minus
  \fontdimen4\font\relax}
\providecommand{\BIBforeignlanguage}[2]{{%
\expandafter\ifx\csname l@#1\endcsname\relax
\typeout{** WARNING: IEEEtran.bst: No hyphenation pattern has been}%
\typeout{** loaded for the language `#1'. Using the pattern for}%
\typeout{** the default language instead.}%
\else
\language=\csname l@#1\endcsname
\fi
#2}}
\providecommand{\BIBdecl}{\relax}
\BIBdecl

\bibitem{dionSurvey}
J.-M. Dion, C.~Commault, and J.~V. der Woude, ``Generic properties and control
  of linear structured systems: a survey.'' \emph{Automatica}, pp. 1125--1144,
  2003.

\bibitem{Reinschke:1988}
K.~J. Reinschke, \emph{{Multivariable control : a graph theoretic approach}},
  ser. Lect. Notes in Control and Information Sciences.\hskip 1em plus 0.5em
  minus 0.4em\relax Springer-Verlag, 1988, vol. 108.

\bibitem{SSCrevisited}
F.~S. Jan Christian~Jarczyk and B.~Alt, ``Strong structural controllability of
  linear systems revisited,'' \emph{Proc. of the 50th IEEE Conference on
  Decision and Control}, pp. 1213--1218, 2011.

\bibitem{Lin_1974}
C.~Lin, ``Structural controllability,'' \emph{IEEE Transactions on Automatic
  Control}, no.~3, pp. 201--208, 1974.

\bibitem{largeScale}
D.~D. Siljak, \emph{Large-Scale Dynamic Systems: Stability and
  Structure}.\hskip 1em plus 0.5em minus 0.4em\relax Dover Publications, 2007.

\bibitem{Murota:2009:MMS:1822520}
K.~Murota, \emph{Matrices and Matroids for Systems Analysis}, 1st~ed.\hskip 1em
  plus 0.5em minus 0.4em\relax Springer Publishing Company, Incorporated, 2009.

\bibitem{reviewPlacement}
P.~Sharon~L. and K.~Rex~K., ``Optimization strategies for sensor and actuator
  placement,'' National Aeronautics and Space Administration Langley Research
  Center, Langley, Virginia 23681, Tech. Rep., 1999.

\bibitem{Khan}
U.~A. Khan and A.~Jadbabaie, ``Coordinated networked estimation strategies
  using structured systems theory,'' \emph{Proc. of the 50th IEEE Conference on
  Decision and Control}, pp. 2112--2117, 2011.

\bibitem{Khan2}
U.~A. Khan and M.~Doostmohammadian, ``A sensor placement and network design
  paradigm for future smart grids,'' \emph{4th IEEE International Workshop on
  Computational Advances in Multi-Sensor Adaptive Processing}, 2011.

\bibitem{Sundaram}
G.~J.~P. M.~Pajic, S.~Sundaram and R.~Mangharam, ``Topological conditions for
  wireless control networks,'' \emph{Proc. of the 50th IEEE Conference on
  Decision and Control}, pp. 2353--2360, 2011.

\bibitem{DBLP:journals/automatica/BoukhobzaH11a}
T.~Boukhobza and F.~Hamelin, ``Observability analysis and sensor location study
  for structured linear systems in descriptor form with unknown inputs,''
  \emph{Automatica}, vol.~47, no.~12, pp. 2678--2683, 2011.

\bibitem{Bullo}
F.~Pasqualetti, A.~Bicchi, and F.~Bullo, ``A graph theoretic characterization
  of power network vulnerabilities,'' in \emph{2011 American Control
  Conference}, San Francisco, CA, USA, June 29 - July 1 2011, pp. 3918 -- 3923.

\bibitem{liu11}
\BIBentryALTinterwordspacing
Y.-Y. Liu, J.-J. Slotine, and A.-L. Barab\'{a}si, ``{Controllability of complex
  networks},'' \emph{Nature}, vol. 473, no. 7346, pp. 167--173, May 2011.
  [Online]. Available: \url{http://dx.doi.org/10.1038/nature10011}
\BIBentrySTDinterwordspacing

\bibitem{Cormen:2001:IA:580470}
T.~H. Cormen, C.~Stein, R.~L. Rivest, and C.~E. Leiserson, \emph{Introduction
  to Algorithms}, 2nd~ed.\hskip 1em plus 0.5em minus 0.4em\relax McGraw-Hill
  Higher Education, 2001.

\bibitem{Shields_Pearson:1976}
R.~W. Shields and J.~B. Pearson, ``Structural controllability of multi-input
  linear systems,'' \emph{IEEE Trans. Autom. Control}, vol. AC-21, no.~3, 1976.

\end{thebibliography}

\end{document}